\documentclass[aps,prl,floatfix,twocolumn,superscriptaddress,showpacs,epsf]{revtex4-1}

\usepackage{amsmath,amssymb,amsfonts}
\usepackage{epstopdf}
\usepackage{epsfig} 
\usepackage{graphicx}
\usepackage{subfigure}
\usepackage{import}
\usepackage{color}
\usepackage{url}

\allowdisplaybreaks

\usepackage[colorlinks=true,linkcolor=blue,citecolor=red]{hyperref}

\newif\ifshowcomments\showcommentstrue

\usepackage{changes}

\makeatletter
\renewcommand\tableofcontents{\@starttoc{toc}}
\makeatother
\newcommand{\figu}[1]
{Fig.~\ref{#1}}

\def\bcen{\begin{center}}
\def\ecen{\end{center}}

             \def\d{\delta}

                    \def\s{\sigma}

\def\G{\Gamma}       \def\D{\Delta}

\def\AA{\buildrel_{\circ}\over{\mathrm{A}}}

  \def\ie{\mbox{\it i.e.\ }}
\def\=={\equiv}

\def\qed{\raise1pt\hbox{\vrule height5pt width5pt depth0pt}}

\def\cG0{{\cal G}_0} 
\def\cG{{\cal G}}

\def\up{\uparrow} \def\down{\downarrow} 

\def\ie{\hbox{\it i.e.\ }} 
\def\ie{\mbox{\it i.e.\ }} \def\=={\equiv}

 \def\ep0{\epsilon_{p}} \def\ed0{\epsilon_{f}}

\def\be{\begin{equation}}
\def\ee{\end{equation}}

\def\cc{c^{\dagger}}
\def\ca{c^{\phantom{\dagger}}}

\newcommand{\braket}[3]{\langle{#1}| {#2} |{#3} \rangle}

\newcommand{\bd}[1]{\mathbf{#1}}

\newcommand{\TNSe}{Ta$_2$NiSe$_5$}

\newcommand{\abinito}{\textit{ab initio}}

\usepackage{fancyhdr}
\begin{document}
\author{Giacomo Mazza}
\thanks{These two authors equally contributed\\
\href{mailto:giacomo.mazza@unige.ch}{giacomo.mazza@unige.ch}\\
\href{mailto:m.roesner@science.ru.nl}{m.roesner@science.ru.nl}}
\affiliation{Department of Quantum Matter Physics, University of Geneva, Quai Ernest-Ansermet 24, 1211 Geneva, Switzerland}
\affiliation{CPHT, CNRS, Ecole Polytechnique, IP Paris, F-91128 Palaiseau, France}
\affiliation{Coll{\`e}ge de France, 11 place Marcelin Berthelot, 75005 Paris, France}
\author{Malte R\"osner}
\thanks{These two authors equally contributed\\
\href{mailto:giacomo.mazza@unige.ch}{giacomo.mazza@unige.ch}\\
\href{mailto:m.roesner@science.ru.nl}{m.roesner@science.ru.nl}}
\affiliation{Radboud University, Institute  for  Molecules and Materials, Heijendaalseweg  135, 6525 AJ Nijmegen, Netherlands}
\author{Lukas Windg\"atter}
\affiliation{Max Planck Institute for the Structure and Dynamics of Matter, Luruper Chaussee 149, 22761 Hamburg, Germany}
\author{Simone Latini}
\affiliation{Max Planck Institute for the Structure and Dynamics of Matter, Luruper Chaussee 149, 22761 Hamburg, Germany}
\author{Hannes H\"ubener}
\affiliation{Max Planck Institute for the Structure and Dynamics of Matter, Luruper Chaussee 149, 22761 Hamburg, Germany}
\author{Andrew J. Millis}
\affiliation{Center for Computational Quantum Physics, Flatiron Institute, New York, NY 10010 USA}
\affiliation{Department of Physics, Columbia University, New York, NY, 10027 USA}
\author{Angel Rubio}
\affiliation{Max Planck Institute for the Structure and Dynamics of Matter, Luruper Chaussee 149, 22761 Hamburg, Germany}
\affiliation{Center for Computational Quantum Physics, Flatiron Institute, New York, NY 10010 USA}
\affiliation{Nano-Bio Spectroscopy Group, Departamento de F\'isica de Materiales, Universidad del Pa\'is Vasco, 20018 San Sebastian, Spain}
\author{Antoine Georges}
\email{ageorges@flatironinstitute.org}
\affiliation{Coll{\`e}ge de France, 11 place Marcelin Berthelot, 75005 Paris, France}
\affiliation{Center for Computational Quantum Physics, Flatiron Institute, New York, NY 10010 USA}
\affiliation{CPHT, CNRS, Ecole Polytechnique, IP Paris, F-91128 Palaiseau, France}
\affiliation{Department of Quantum Matter Physics, University of Geneva, Quai Ernest-Ansermet 24, 1211 Geneva, Switzerland}
%
\title{Nature of symmetry breaking at the excitonic insulator transition: \TNSe}
\begin{abstract}
\TNSe~is one of the most promising materials for hosting  an excitonic insulator ground state. 
While a number of experimental observations have been interpreted in this way, the precise nature of the symmetry breaking occurring in \TNSe, 
the electronic order parameter, and a realistic microscopic description of the transition mechanism are, however, missing. 
By a symmetry analysis based on first-principles calculations, we uncover the \emph{discrete} lattice symmetries which are broken at the transition.
We identify a purely electronic  order parameter of excitonic nature that breaks these discrete crystal symmetries and 
contributes to the experimentally observed lattice distortion from an orthorombic to a monoclinic phase.
Our results provide a theoretical framework to understand and analyze the excitonic transition in \TNSe~and settle the fundamental questions about symmetry breaking governing the spontaneous formation of excitonic insulating phases in solid-state materials.
\end{abstract}

\maketitle

\paragraph{Introduction.}
Spontaneous symmetry breaking is a fundamental organizing principle for understanding the emergence of long-range order.
Identifying the underlying symmetry breaking is thus a key step in the characterization of the ordered phase. 
This can be an elusive task when the symmetry breaking field cannot be directly tuned experimentally or when different types of ordering are coupled.
The so-called {\it excitonic insulator} is a prominent example of such an elusive state of matter.
This phase~\cite{kozlov_divalent_crystal, keldysh_collective_properties_excitons, kohn_excitonic_ins, halperin_rice_rmp} has been identified with the spontaneous condensation of excitons (bound electron-hole pairs) stemming from the Coulomb attraction between electrons and holes in the conduction and valence bands.
Excitonic condensation has been observed and intensively investigated in specially designed devices such as bilayer quantum Hall systems~\cite{eisenstein_review_2014, kellogg_double_electron_gas_prl2004, spielman_double_quantum_hall_ferromagnet_prl2000, Li_dblg_nat_phys_2017, Liu_quantum_hall_drag_NatPhys2017} or by photo stimulation of electron-hole pairs~\cite{snoke_review_2002, Butov_exciton_condensation_traps, butov_exciton_condensate_Nat2002}.
In contrast, spontaneous excitonic condensation in bulk materials still remains an open question and its detection a major challenge. 

\TNSe~has been proposed as a candidate material hosting a homogeneous excitonic condensate~\cite{disalvo_Ta2NiSe5,wakisaka_photoemission_TNS,seki_TNS,exp_ei_Ta2NiSe5}, 
\ie without charge or other non-zero momentum order~\cite{di_salvo_TiSe2,exp_ei_TiSe2,kogar_TiSe2}.
\TNSe~undergoes a structural transition from a high-temperature orthorhombic to a low-temperature monoclinic phase 
at $T_s \simeq 328~\mathrm{K}$~\cite{disalvo_Ta2NiSe5,nakano_antiferro_electric_distorsion,TNS_ultrathin_film}.
Proposed evidence for excitonic condensation occurring simultaneously with the structural transition includes a characteristic flattening of the valence band close to the $\G$ point~\cite{wakisaka_photoemission_TNS,seki_TNS,mor_ufast_ei}, the opening of a gap in the electronic spectrum~\cite{PhysRevB.99.075408,exp_ei_Ta2NiSe5,TNS_fano_resonance}, and coherent oscillations reminiscent of the excitation of an amplitude mode of the condensate~\cite{kaiser_TNS}.
Due to its characteristic chain structure \TNSe~has so far been interpreted as a quasi one-dimensional excitonic insulator~\cite{sugimoto_strong_couplingTNS,kaneko_ortho_to_mono} and Kaneko {\it et al.}~\cite{kaneko_ortho_to_mono} proposed a scenario in which the coupling of the excitonic condensate with phonons gives rise to a combined excitonic and structural instability.

The following symmetry considerations, however, call into question the very notion of 
excitonic condensation in the solid-state context.
Condensation implies the breaking of a \emph{continuous} symmetry.
In the case of excitonic condensation this would be the breaking of the $U_X(1)$ symmetry related to the
conservation of \emph{relative} charge between valence and conduction states.
Bulk materials, however, generally lack such a $U_X(1)$ symmetry due 
to the hybridization between conduction and valence bands.
The only continuous symmetry being present is the one related to the 
\emph{global} charge conservation $U_N(1)$.
Nonetheless, internal {\it discrete} symmetries of the solid, such as crystal 
symmetries, can result in an approximate realization of the relative
charge conservation 
with symmetry-forbidden hybridizations in particular regions of the Brillouin zone.
Therefore, we propose here that the spontaneous hybridization 
introduced by an excitonic instability 
represents a general mechanism for breaking internal discrete symmetries rather than 
a condensation phenomenon resulting from the breaking of a continuous symmetry. 

We demonstrate this concept for the case of \TNSe~by uncovering the symmetries that are broken by an excitonic instability.
To this end, we construct a minimal yet realistic model for \TNSe~including its electronic band structure and electron-electron interactions from first principles. 
We show the existence of an electronic instability of excitonic origin leading to an electronic phase that breaks 
a set of discrete symmetries of the high-temperature orthorombic phase and is compatible 
with the low-temperature monoclinic structure.
This analysis settles the fundamental question of identifying which symmetries are broken at the excitonic transition 
and is therefore of general relevance to the understanding and the eventual control of such transitions 
in \TNSe~and solid-state materials in general.

\paragraph{Crystal symmetries in \TNSe.}

\begin{figure}[t]
	\includegraphics[width=0.92\columnwidth]{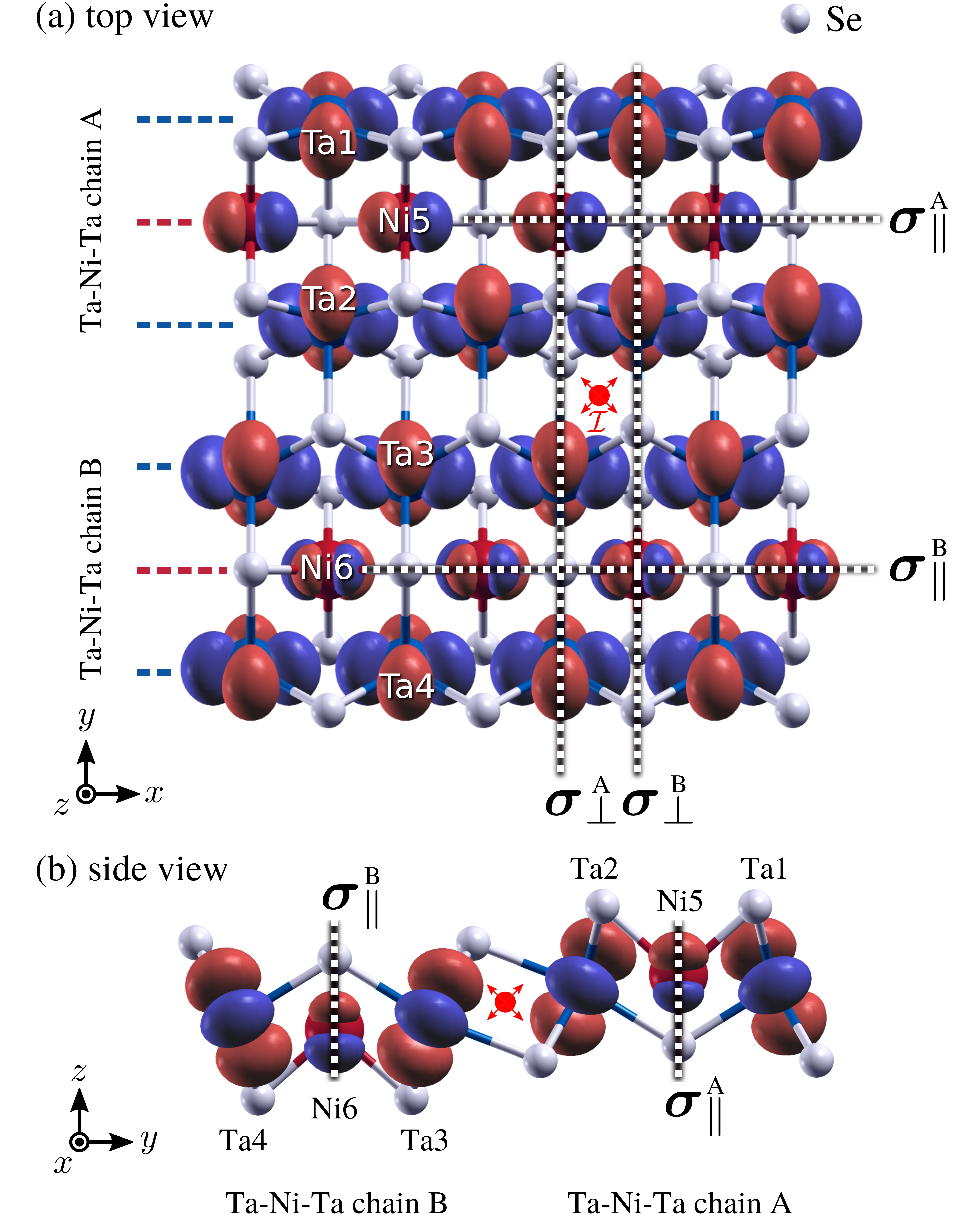}
	\caption{Top and side view of the \TNSe\ lattice structure including isosurfaces of Wannier wave functions localized at Ta and Ni positions (red/blue correspond to opposite MLFW-amplitude signs). Dashed lines (red dots) indicate reflection (inversion) symmetries.        
	}
	\label{fig1}
\end{figure}

\begin{figure}[ht]
	\includegraphics[width=0.99\columnwidth]{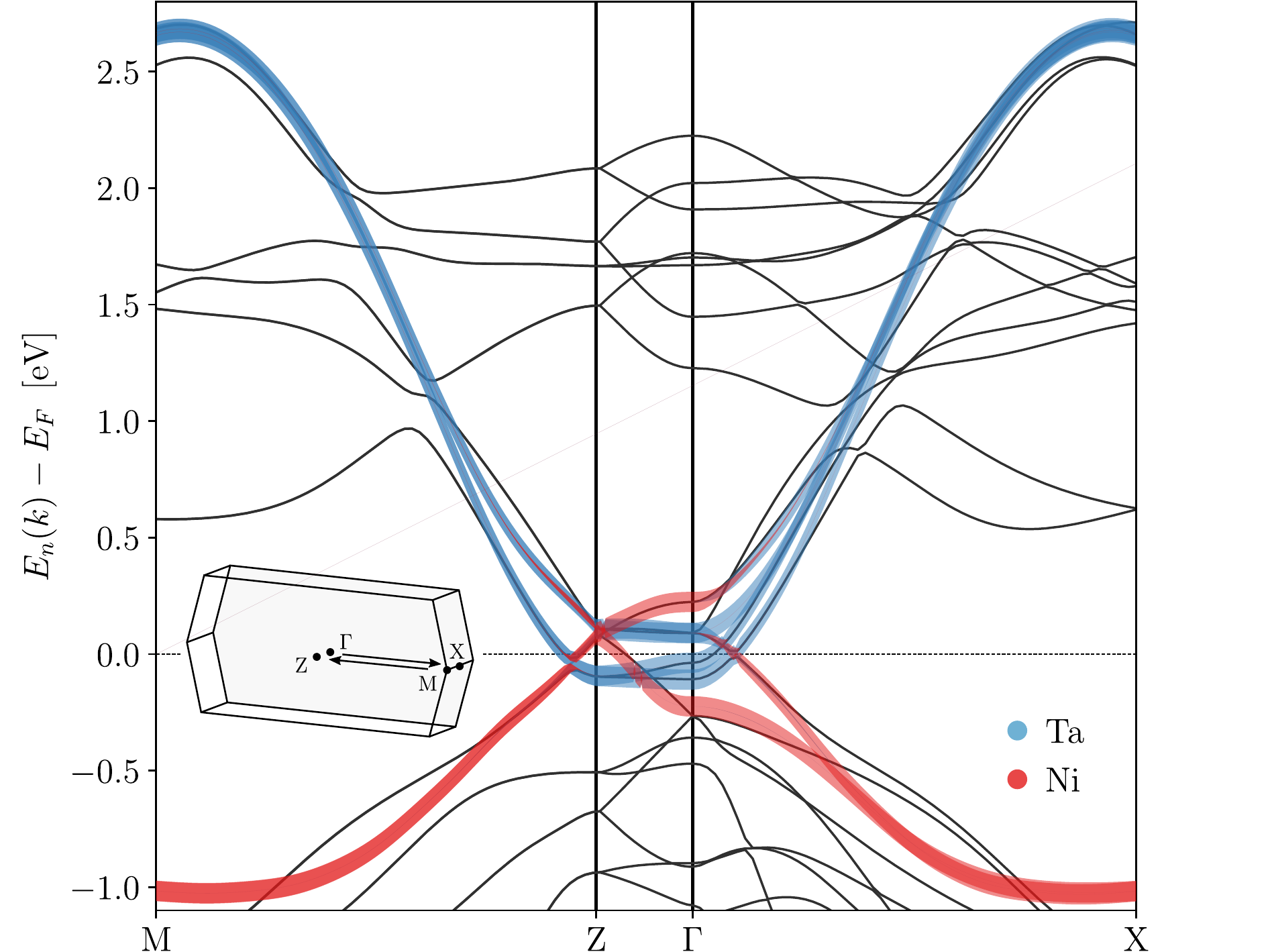}
	\caption{Ab initio band structure (black lines) together with a fat-bands representation of the Wannier model.
          Thick blue (red) lines represent Ta (Ni) contributions as resulting from the Wannier model.
	}
	\label{fig2}
\end{figure}

We perform DFT calculations in the high-temperature orthorhombic phase, with $a \simeq 3.51 \AA$ and $b \simeq 15.79 \AA$ being
the lattice constants in $x$- and $y$-directions of the Bravais lattice~\cite{suppl}.
This unit-cell is composed of two formula units with atoms arranged in two parallel Ta-Ni-Ta 
chains (A/B) along the $x$-direction. 
The chains are shifted by half a lattice constant along $x$ and displaced along $z$, which results in four reflection symmetries with planes parallel and perpendicular to the Ta-Ni-Ta chains ($\boldsymbol{\s}_{\parallel/\perp}^{A/B}$) and one inversion symmetry point $\bd{\cal I}$, as depicted in \figu{fig1}.
Based on these \abinito\ calculations, we construct six d$_\text{xz}$-like maximally localized Wannier functions (MLWF) centered at the Ta and Ni positions, which are shown in \figu{fig1}. 
In each Ta-Ni-Ta chain the Ta-centered d$_\text{xz}$ orbitals, $\varphi_{\text{Ta}}(\vec{R})$, are aligned along the chains and tilted around the $x$-axis following the Ta-Se bonds [see \figu{fig1}(b)]. 
The Ni-centered d$_\text{xz}$ MLWFs, $\varphi_{\text{Ni}}(\vec{R})$, are also parallel to the Ta-Ni-Ta chains, but rotated by $45^\circ$ around the $y$-axis. 
The Se contributions are thus indirectly accounted for by deforming and 
rotating the d$_\text{xz}$ orbitals.

The reflection symmetries act differently on the Ta and Ni-centered MLWFs. 
While the $\varphi_{\text{Ta}}(\vec{R})$ MLWFs are unaffected by all 
reflections,  $\varphi_{\text{Ni}}(\vec{R})$ change sign under $\boldsymbol{\s}_\perp$.
It follows that the intra-chain Ta-Ta and Ni-Ni hoppings $t_\text{TaTa/NiNi}(\vec{R}) = \braket{\varphi_\text{Ta/Ni}^{A/B}(\vec{R})}{\hat{H}}{\varphi_\text{Ta/Ni}^{A/B}(0)}$ have opposite signs. 
We find $t(\vec{0})_{\text{TaTa}} \approx -640\,$meV and $t(\vec{0})_{\text{NiNi}} \approx 250\,$meV, which are respectively mainly responsible for the conduction and valence bands dispersions in the $\overline{MZ}$ and $\overline{\G X}$ directions of the Brillouin zone.
These bands are  about $2.5\,$eV and $1.5\,$eV wide with predominant Ta (blue) and Ni (red) character as visible in the Wannier-interpolated band structure in ~\figu{fig2}.
Conduction and valence bands overlap along $\overline{Z\G}$ where the bands become much less dispersive and are mainly characterized by bonding/anti-bonding splittings of the Ta- and Ni- states between the two Ta-Ni-Ta chains.

Importantly, hopping matrix elements between Ta- and Ni-like states within 
the chains are \emph{not} forbidden by any symmetry.
In fact, even in the orthorhombic phase 
we obtain non-zero matrix elements between Ta- and Ni-MLWFs 
within the same chain $t_{\text{Ta} \text{Ni}}(\vec{R})$. 
Specifically, $t_\text{TaNi}(\vec{0}) \approx 36\,$meV which decreases with the distance $R_x$ along the chains. 
This result is universal to all tested DFT exchange-correlation 
functionals~\cite{suppl} and shows that this kind of Ta-Ni 
hybridization cannot spontaneously form due to exciton condensation 
below a critical temperature~\cite{kaneko_ortho_to_mono, ejima_extended_falikov_kimball, sugimoto_strong_couplingTNS}.

In contrast to that, we will show below that the excitonic instability can break the crystal 
symmetries that constrain Ta-Ni hybridization in the high-temperature phase.
In particular, the reflection symmetries $\boldsymbol{\sigma}^{A/B}_\perp$ 
constrain  the Ta-Ni hoppings to change sign under $\boldsymbol{\sigma}_\perp^{A/B}$
implying that the Wannier Hamiltonian averaged along the $x$-direction 
is block-diagonal with respect to the Ta- and Ni- states
\begin{equation}
\hat{H}(k_x=0,R_y) \equiv 
\sum_{R_x} 
\hat{H}(R_x,R_y) 
=
\left(
\begin{matrix}
\hat{h}_\text{Ta}(R_y) & \hat{0}\\
\hat{0} & \hat{h}_\text{Ni}(R_y)
\end{matrix}
\right).
\label{eq:H_block_diag}
\end{equation}
In momentum space, $t_{\text{Ta} \text{Ni} }(k_x=0, k_y) = 0$
so that the bands along the $\overline{Z\G}$ path have purely Ta or Ni character,  \figu{fig2}.
Therefore, any excitonic instability resulting from a spontaneous Ta-Ni hybridization must 
break the $\boldsymbol{\sigma}^{A/B}_\perp$ symmetry and show-up along the 
$\overline{Z\G}$ direction.
We provide evidence of such an instability by considering the effect of the electron interactions
in a minimal model derived from the above symmetry analysis.

\paragraph{Minimal Model.}  
We consider a two-dimensional lattice with six atoms per unit cell. The electronic Hamiltonian 
\begin{equation}  
  \hat{\cal H} = \hat{H}_{hop} + \hat{H}_{U} + \hat{H}_{V}
  \label{eq:Htoy}
\end{equation}
includes a hopping term $\hat{H}_{hop}$, a local $\hat{H}_{U}$ Coulomb interaction term, and nearest-neighbor $\hat{H}_V$ one.
We define $\Psi_{\vec{R}\s}^{\dagger} \equiv
\left(
\begin{matrix}
\cc_{1 \s}(\vec{R}) \ldots \cc_{6 \s}(\vec{R})
\end{matrix}  \right)$,
where $\cc_{j \s}(\vec{R})$
creates an electron with spin $\s$ in a localized orbital on the $j$-th atom (labels in ~\figu{fig1}) 
of the unit cell $\vec{R}$.
\begin{equation}
\hat{H}_{hop} = \sum_{\vec{R} \s} \sum_{\vec{\d}} \Psi_{\vec{R}+\vec{\d}\s}^{\dagger} \mathbf{T}(\vec{\d})  
\Psi_{\vec{R}\s}
\end{equation}
contains intra-cell, $\mathbf{T}(\vec{0})$, as well as inter-cell, $\mathbf{T}(\pm a, \pm b)$, terms. 
The matrix elements are chosen consistently with the above symmetry requirement
and in order to reproduce the main features of the Wannier band structure~\cite{suppl}. 

\begin{figure}
	\includegraphics[width=0.99\columnwidth]{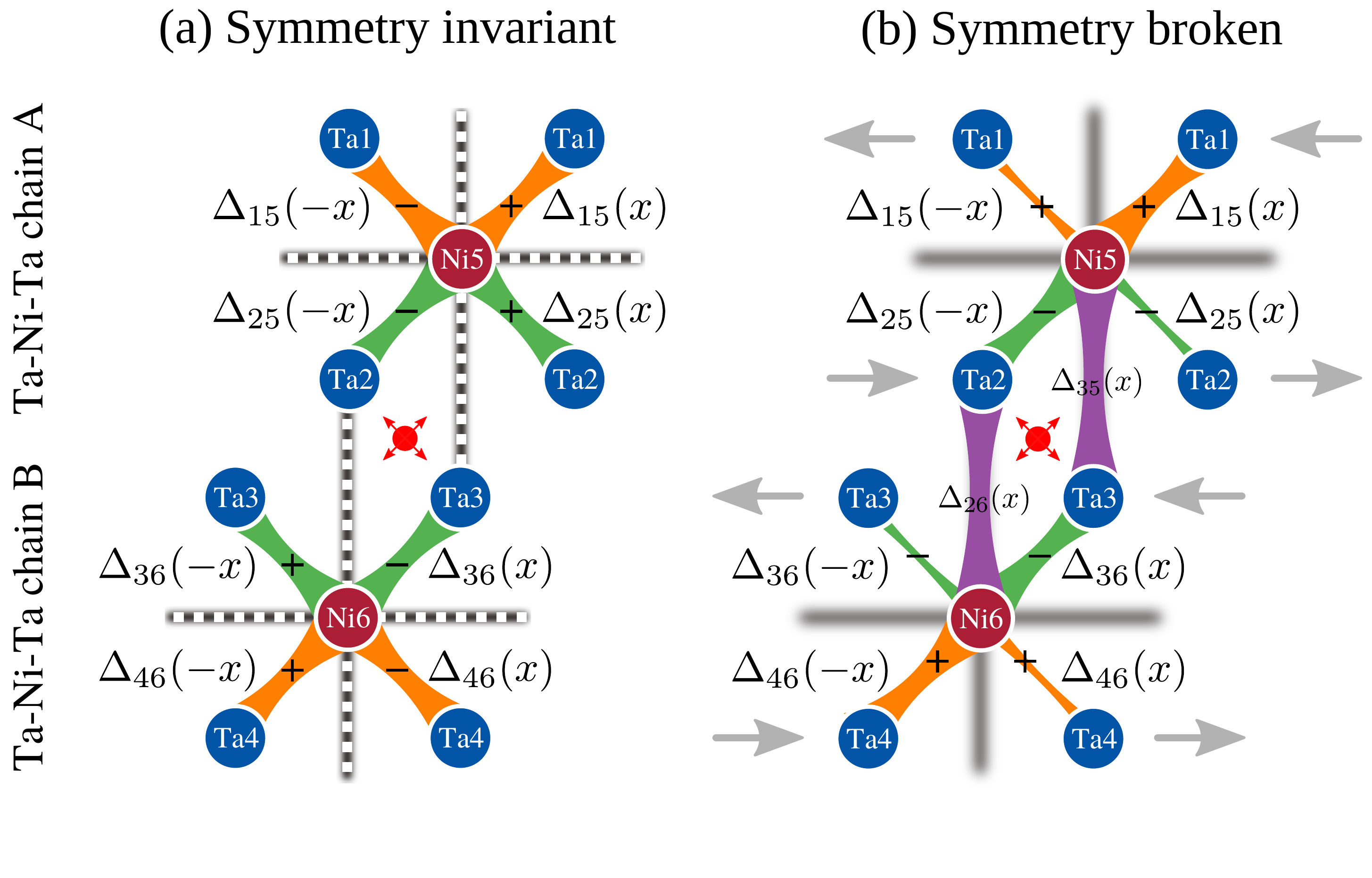}
	\caption{Scheme of nearest neighbor Ta-Ni hybridization in 
		the symmetry invariant (a) and symmetry-broken (b) phases. 
		Thickness of the lines connecting the atoms indicates the absolute value of the hybridization.}
	\label{fig4}
\end{figure}

\begin{figure*}
	\includegraphics[width=0.95\linewidth]{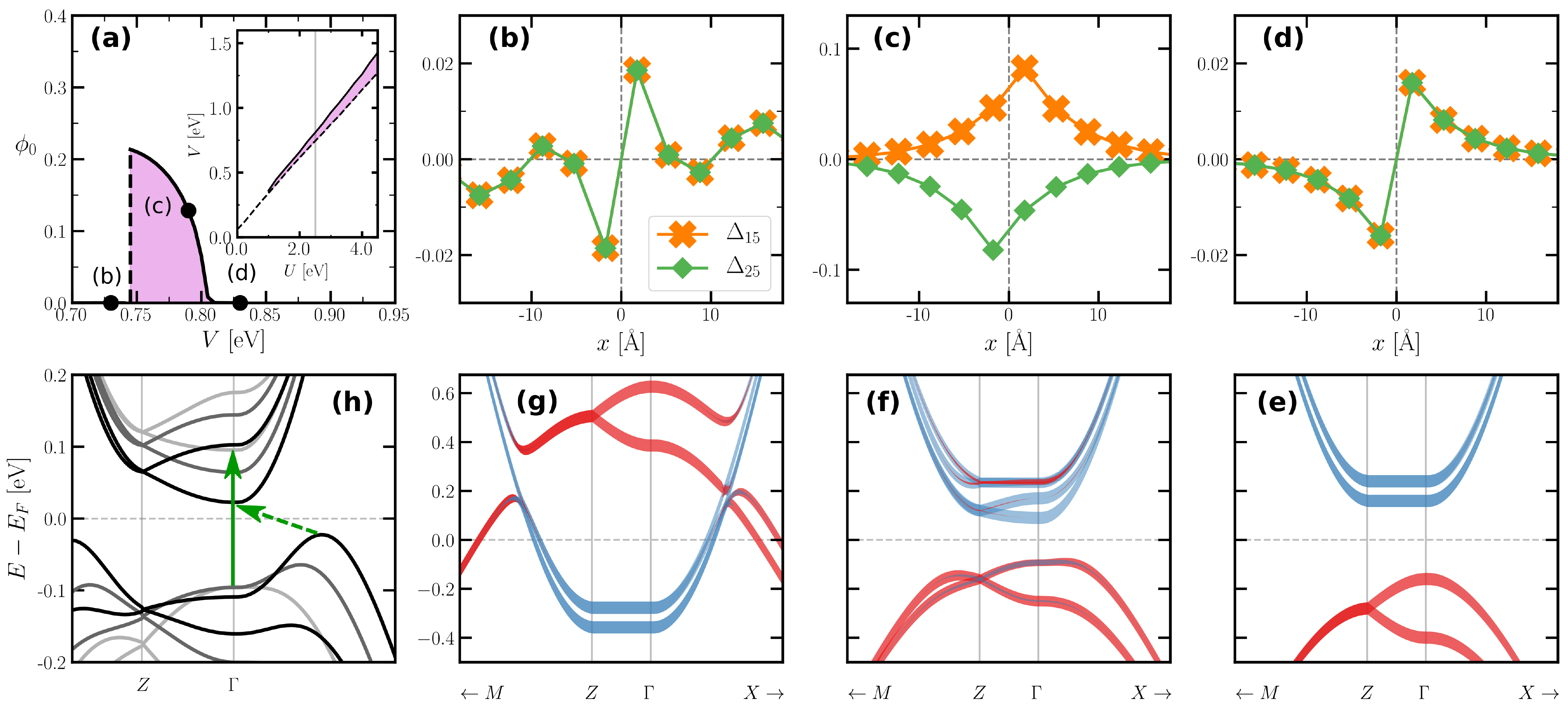}
	\caption{(a) Order parameter at zero-temperature as function of $V$ for $U=2.50~$eV.
	  Dots represent $V=0.73$, $V=0.785$ and $V=0.83$ corresponding to panels (b)-(d).
	  Inset: phase diagram in $U-V$ plane. Shaded region corresponds to the symmetry-broken phase.
	  Dashed line indicates a metal-insulator Lifshitz transition. 
	  (b)-(d): Ta-Ni hybridization along Ta-Ni-Ta A chain
	  in the symmetric (b) and (d) and symmetry-broken (c) phases.
	  Crosses/diamonds correspond to upper/lower part of the Ta-Ni-Ta chain.
	  (e)-(g): Bands along $M-Z-\G-X$ corresponding to panels (b)-(d). 
	  Red/blue corresponds to Ni/Ta character.
	  (h): Gap evolution inside the broken-symmetry phase. $U=2.50~$eV and 
	  $V=0.79,0.77,0.75~$eV from light grey to black lines. 
	  Full/dashed green arrows highlight the direct/indirect gap.
	}
	\label{fig3}
\end{figure*}
The electrons interact through a local Hubbard-like term
\begin{equation}
\hat{H}_U = U \sum_{\vec{R}} \sum_{j} 
\hat{n}_{j \up}(\vec{R}) \hat{n}_{j \down}(\vec{R}),
\end{equation}
where we assumed the same $U$ for the six atoms, 
as supported by a constrained RPA~\cite{PhysRevB.70.195104} analysis of the Coulomb matrix elements 
($U_\text{Ta} \approx 2.1\,$eV, $U_\text{Ni} \approx 2.4\,$eV).
The next leading terms are intra-chain density-density interactions
between neighbouring Ta and Ni atoms 
($V \approx 0.9\,$eV) 
\begin{equation}
\begin{split}
\hat{H}_V =& V \sum_{j=1,2} \sum_{\vec{R}  \s \s'} \left[ \hat{n}_{j \s}(\vec{R}) +    
\hat{n}_{j \s}(\vec{R}+\vec{\d}_x) 
\right] \hat{n}_{5 \s'}(\vec{R})  \\
+&
V \sum_{j=3,4} \sum_{\vec{R}  \s \s'} \left[ \hat{n}_{j \s}(\vec{R}) +    
\hat{n}_{j \s}(\vec{R}-\vec{\d}_x) 
\right] \hat{n}_{6 \s'}(\vec{R}). 
\end{split}
\end{equation}
The symmetries of the Hamiltonian are revealed by an investigation of the intra-chain Ta-Ni hybridization as a function of the distance along $x$ between the Ta and Ni atoms
\begin{equation}
  \D_{ij}(x) = \langle \underbrace{\cc_i(R_x,0)}_{\text{Ta site}} \underbrace{\ca_j(0,0)}_{\text{Ni site}}\rangle,
  \label{eq:def_delta}
\end{equation}
where $i=1,2(3,4)$ and $j=5(6)$ label the Ta and Ni states, respectively, for the A(B) chain. 
For each chain, $x$ is defined by taking the Ni atom in that chain as origin, so that  $x = R_x \mp a/2$ ($-$ for A and $+$ for B).
We have dropped the spin index as we focus on the spin-singlet case.
For the A-chain, $\D_{15}(x)$ and $\D_{25}(x)$, i.e. the hybridizations between the lower and upper Ta with the central Ni state of the A-chain, transform as
\begin{equation}
\boldsymbol{\s}^A_\perp \D_{15}(x) = -\D_{15}(-x)\qquad
\boldsymbol{\s}^A_\parallel \D_{15}(x) = \D_{25}(x)
\label{eq:delta_symmetry}
\end{equation}
so that $\D_{15}(x)=-\D_{15}(-x)$ and $\D_{15}(x) = \D_{25}(x)$, as depicted in Fig.~\ref{fig4}(a).
Similarly, reflection symmetries for the B-chains 
imply $\D_{36}(x) = -\D_{36}(-x)$ and $\D_{46}(x) = -\D_{46}(-x)$.

We investigate the possible breaking of the above symmetries due to electronic interactions, by utilizing a Hartree-Fock (HF) variational wave-function allowing for a spatially homogeneous order parameter of the form: 
\begin{align}
	\vec{\phi} =
		\begin{pmatrix}
			\phi_{15} \\
			\phi_{25} \\
			\phi_{36} \\
			\phi_{46} 
		\end{pmatrix}
		\,\,\,,\,\,\,\phi_{ij} \equiv  \D_{ij}(a/2) +\D_{ij}(-a/2).
\end{align}
The four $\phi_{ij}$ are in general independent, allowing in principle for 16 different phases
corresponding to the different breaking patterns of the reflections and inversion symmetries.
Here, we focus on the symmetry breaking channel consistent with the low-temperature monoclinic phase of \TNSe. 
In the monoclinic phase all reflections  are broken,
while their products $I^{A/B} = \boldsymbol{\s}^{A/B}_\perp \boldsymbol{\s}_\parallel^{A/B}$ and 
the inversion ${\cal I} = I^{A/B} {\cal T}$ 
(${\cal T}$ being the translation between the two Ni atoms) are preserved.
This constrains the components $\phi_{ij}$ as $\phi_{15} = -\phi_{25}$  and  $\phi_{36}=-\phi_{46}$, 
due to preservation of $I^{A/B}$ and  $\phi_{15} = \phi_{46}$ and $\phi_{25} = \phi_{36}$ due to $\cal I$,
leading to an order parameter of the form $\vec{\phi}=\phi_0(+1,-1,-1,+1)$.

The obtained zero-temperature phase diagram in the $U$-$V$ plane, \figu{fig3}(a), shows three distinct regions. 
At fixed value of $U$, the order parameter $\phi_0$ vanishes for $V$ smaller than a lower critical value [$V<V_l^*(U)$] 
and for $V$ larger than an upper critical value [$V>V_u^*(U)$].
In these regions the  Ta-Ni hybridizations transform in accordance with Eq.~(\ref{eq:delta_symmetry}),
as shown in \figu{fig3}(b) and (d). 
These two symmetric ground states are characterized by different electronic properties.
For $V < V_l^*(U)$ [\figu{fig3}(b) and (g)] the valence and conduction bands overlap,
while for $V > V_u^*(U)$ [\figu{fig3}(d) and (e)] the bands are separated by an energy gap.
%

In the intermediate region $V_l^*(U) < V < V_u^*(U)$ [\figu{fig3}(c) and (f)] a solution with $\phi_0 \neq 0$ is stabilized. 
This is the hallmark of the excitonic instability as witnessed in \figu{fig3}(f) by the emergence of a sizable hybridization between valence and conduction bands all along the 
$\overline{Z\G}$ path.
Valence and conduction bands acquire a strong Ta and Ni character respectively (which is absent in the symmetric phase) and the degeneracy of the Ta-like conduction bands along $\overline{Z\G}$ is lifted by hybridization with Ni-like valence bands.
In real space this translates into broken $\boldsymbol{\sigma}^{A/B}_\perp$ and $\boldsymbol{\sigma}^{A/B}_\parallel$ symmetries yielding finite $\D_{35}$ and $\D_{26}$, which couple the two chains, \figu{fig4}(b).

The upper valence band develops a mostly flat dispersion around $\Gamma$.
While this has so far  been interpreted as a distinctive signature of an homogeneous 
excitonic condensate, we show here that the interpretation is not unique.
In fact this feature is a result of a direct-to-indirect gap insulator transition,
driven by the splitting between the hybridized bands along $\overline{Z\G}$,
that can occur inside the broken-symmetry phase [\figu{fig3}(h)].
By decreasing $V$ the splitting increases, while the bottom of the conduction band 
moves closer to the Fermi level. At $V=V_l^*(U)$ the conduction band crosses the Fermi level and
the system undergoes a Lifshitz transition accompanied by the formation of a Fermi surface in the metallic phase.
We find here that this is a first-order transition which restores the symmetry for $V < V_l^*(U)$. 
If we allowed for a non-homogeneous $\vec{\phi}(\vec{R})$
order-parameter a finite momentum instability could occur near this 
point~\cite{domon_fflo}.
The symmetry-broken phase is found only in a small region of the phase space close to the Lifshitz transition. 
In this regime the symmetric phase is characterized by a very small gap, 
reinforcing the relevance of the above phase transition for \TNSe~which in the high-temperature 
phase has been reported to be a zero-gap semiconductor~\cite{exp_ei_Ta2NiSe5}. 
The symmetry-broken region shrinks as the Hubbard $U$ is decreased until
it disappears for $U \lesssim 1.25~\mathrm{eV}$ for which $V_l^*(U)$ and $V_u^*(U)$ merge into the Lifshitz transition line. 
We highlight that our constrained RPA values for $U$ and $V$ are in close vicinity of the symmetry-broken region.

\paragraph{Structural phase transition.}
The electronic configuration associated with the excitonic phase 
 is not compatible with the symmetries of the lattice.
 This implies that the electronic order parameter must have a linear
 coupling to lattice modes breaking the crystal symmetry~\cite{kaneko_ortho_to_mono,PhysRevB.90.195118}. 
 Hence, the excitonic transition will co-exist with a structural transition,
 as indeed observed experimentally.
 From the pattern of the Ta-Ni hybridizations in the broken-symmetry phase [\figu{fig4}(b)], one anticipates a distortion  
 of the unit cell in which Ta atoms from the same Ta-Ni-Ta chain are tilted in opposite directions  [arrows in \figu{fig4} (b)].
 This corresponds to a structural transition from the orthorhombic
 to the monoclinic structure, which we confirm to be present by performing a full structural relaxation within DFT~\cite{suppl}. 
 The interplay between the electronic and lattice instability is an interesting question for future investigations~\cite{in_prep}.

\paragraph{Conclusions.} 

We have performed a symmetry analysis backed up by first-principle calculations of \TNSe, 
with general implications for the excitonic transitions in solids. 
While this transition has been so far understood as
a condensation phenomenon resulting from a continuous $U_X(1)$
symmetry breaking~\citep{kaneko_ortho_to_mono}
we show that in realistic solids there is no such symmetry: 
the purely electronic transition 
corresponds to the breaking of discrete symmetries only. 
Important consequences include that all collective modes are gapped and that there 
is no dissipationless transport or excitonic superfluidity.  
%
We identify explicitly all discrete symmetries relevant for the structural phase transition in
 \TNSe, including the corresponding electronic order parameters and provide clear evidence for a transition into an excitonic insulator phase. 
This transition breaks these symmetries in a manner consistent with the experimentally observed lattice distortion into a monoclinic phase, and the order parameter couples linearly to lattice modes. 

%
Because we find a spontaneous electronic instability for realistic values of the interactions, 
our results suggest an electronic contribution to the coupled 
transition~\cite{watson2019band,subedi2020orthorhombictomonoclinic,tang2020noncoulomb}.
However, a definitive confirmation of this point calls for experimental probes which can selectively
address the electronic and lattice degrees of freedom. 
In the context of iron-based superconductors, where a similar question arises in relation to nematicity,
it has proven possible to probe the electronic component of the susceptibility associated with the
nematic instability~\cite{chu_fisher_nematic_susceptibility, massat_gallais_charge_nematicity}.
Ultra-fast spectroscopies offer another possible route~\cite{kaiser_TNS,mor_ufast_ei,PhysRevLett.119.247601,PhysRevLett.119.247601,Non-thermal_Porer_2014}, by exploiting the very different time scales associated with electronic and lattice degrees of freedom.

\begin{acknowledgments}
Discussions with Jennifer Cano, Denis Gole\v z, Edoardo Baldini, Selene Mor, Tatsuya Kaneko, and Jernej Mravlje are gratefully acknowledged.  
We thank Merzuk Kaltak for sharing his (c)RPA implementation \cite{kaltak_merging_2015} with us.
This work was supported by (AG, GM) the European Research Council (ERC-319286-QMAC) and (AM) the 
US Department of Energy under grant DE-SC 0019443.
GM acknowledges support from the Swiss National Science Foundation Ambizione grant PZ00P2\_186146.
SL, LW, HH and AR were supported by the European Research Council (ERC-2015-AdG694097),
the Cluster of Excellence AIM, Grupos Consolidados (IT1249-19) and SFB925. 
SL acknowledges support from the Alexander von Humboldt foundation. 
The Flatiron Institute is a division of the Simons Foundation.
\end{acknowledgments}

%

\end{document}


\author{Giacomo Mazza}
\thanks{These two authors equally contributed\\
\href{mailto:giacomo.mazza@unige.ch}{giacomo.mazza@unige.ch}\\
\href{mailto:m.roesner@science.ru.nl}{m.roesner@science.ru.nl}}
\affiliation{Department of Quantum Matter Physics, University of Geneva, Quai Ernest-Ansermet 24, 1211 Geneva, Switzerland}
\affiliation{CPHT, CNRS, Ecole Polytechnique, IP Paris, F-91128 Palaiseau, France}
\affiliation{Coll{\`e}ge de France, 11 place Marcelin Berthelot, 75005 Paris, France}
\author{Malte R\"osner}
\thanks{These two authors equally contributed\\
\href{mailto:giacomo.mazza@unige.ch}{giacomo.mazza@unige.ch}\\
\href{mailto:m.roesner@science.ru.nl}{m.roesner@science.ru.nl}}
\affiliation{Radboud University, Institute  for  Molecules and Materials, Heijendaalseweg  135, 6525 AJ Nijmegen, Netherlands}
\author{Lukas Windg\"atter}
\affiliation{Max Planck Institute for the Structure and Dynamics of Matter, Luruper Chaussee 149, 22761 Hamburg, Germany}
\author{Simone Latini}
\affiliation{Max Planck Institute for the Structure and Dynamics of Matter, Luruper Chaussee 149, 22761 Hamburg, Germany}
\author{Hannes H\"ubener}
\affiliation{Max Planck Institute for the Structure and Dynamics of Matter, Luruper Chaussee 149, 22761 Hamburg, Germany}
\author{Andrew J. Millis}
\affiliation{Center for Computational Quantum Physics, Flatiron Institute, New York, NY 10010 USA}
\affiliation{Department of Physics, Columbia University, New York, NY, 10027 USA}
\author{Angel Rubio}
\affiliation{Max Planck Institute for the Structure and Dynamics of Matter, Luruper Chaussee 149, 22761 Hamburg, Germany}
\affiliation{Center for Computational Quantum Physics, Flatiron Institute, New York, NY 10010 USA}
\affiliation{Nano-Bio Spectroscopy Group, Departamento de F\'isica de Materiales, Universidad del Pa\'is Vasco, 20018 San Sebastian, Spain}
\author{Antoine Georges}
\email{ageorges@flatironinstitute.org}
\affiliation{Coll{\`e}ge de France, 11 place Marcelin Berthelot, 75005 Paris, France}
\affiliation{Center for Computational Quantum Physics, Flatiron Institute, New York, NY 10010 USA}
\affiliation{CPHT, CNRS, Ecole Polytechnique, IP Paris, F-91128 Palaiseau, France}
\affiliation{DQMP, Universit{\'e} de Gen{\`e}ve, 24 quai Ernest Ansermet, CH-1211 Gen{\`e}ve, Suisse}

\title{Supplemental information for:\\
Nature of symmetry breaking at the excitonic insulator transition: Ta$_2$NiSe$_5$}

\maketitle
\onecolumngrid

\section{Ab-initio Calculations}

Our ab-initio calculations are performed using density functional theory (DFT) initially applying the generalized gradient approximation (GGA / PBE) \cite{perdew_generalized_1996} within the PAW formalism \cite{blochl_projector_1994} as implemented in the Vienna Ab initio Simulation Package (VASP) \cite{kresse_efficiency_1996,kresse_efficient_1996}. 
We start with fully relaxing the internal atomic coordinates of an orthorhombic unit cell with $a \approx 3.51\,\AA$, $b \approx 14.07\,\AA$, and $c \approx 15.79\,\AA$ as lattice constants in reasonable agreement with experimental value\cite{doi:10.1021/ic00216a027}. To this end we use a $12 \times 12 \times 3$ $K$-grid and an energy cut-off of $368\,$eV. The positions are optimized until all forces are smaller than $0.005\,$eV$/\AA$.

Due to the layered structure, screening is reduced so that enhanced Coulomb interactions are expected. To take the resulting correlation effects into account, we use the modified Becke-Johnson exchange potential~\cite{becke_mbj}, which has been shown to have a similar accuracy as hybrid functional or $GW$ approaches~\cite{PhysRevLett.102.226401}. The involved $c_\text{mbj}$ parameter is self-consistently find to be $c_\text{mbj} = 1.26$ on a $20 \times 20 \times 5$ $K$-grid.

The resulting Kohn-Sham states are subsequently projected onto six $d_\text{xz}$-like Wannier orbitals centered at the Ta and Ni sites, which are maximally localized using the Wannier90 package\cite{mostofi_wannier90:_2008} applying an inner (frozen) window of about $\pm0.3\,$eV around the Fermi level. Thereby, the overlap between the original Kohn-Sham states and the reconstructed ones is maximized throughout the low-energy window.

These six maximally localized Wannier functions are also used as the basis for the evaluation of the Coulomb matrix elements calculated within the constrained Random Phase Approximation (cRPA) \cite{PhysRevB.70.195104} as recently implemented by M. Kaltak within VASP\cite{kaltak_merging_2015}. We use in total $120$ bands (about $50$ unoccupied) and apply the weighted disentanglement procedure from Ref.\,\onlinecite{PhysRevB.83.121101}.

\section{Structural Phase transition}
 
\begin{figure}
	\centering
	\includegraphics[width=0.75\textwidth]{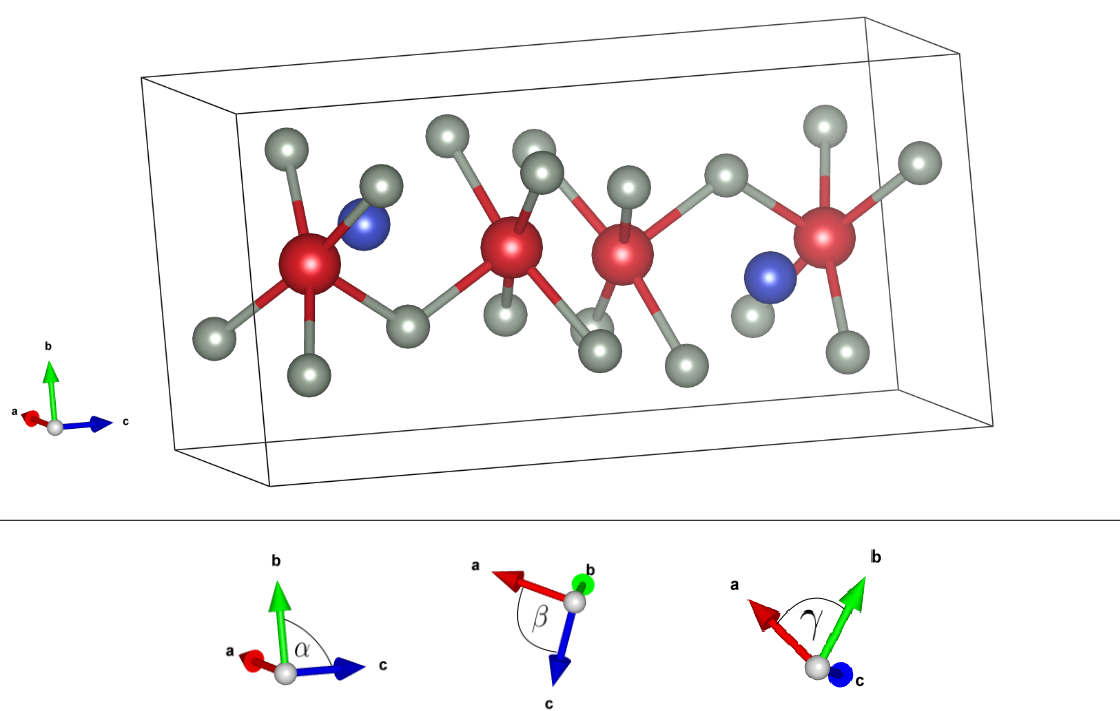}
	\caption{Orthorhombic unit cell. The Tantalum atoms are depicted in red, the Nickel atoms in blue and the Selenium atoms in grey. In the lower part the three angles of the cell are displayed}
	\label{fig:ortho_unit_cell}
\end{figure}

Starting from the relaxed orthorhombic geometry we introduce a small distortion to $\beta = 90^\circ + \delta$ (see figure \ref{fig:ortho_unit_cell}) to seed the monoclinic phase and perform a full relaxation allowing for an optimization of the cell shape, cell-volume and atomic coordinates afterwards. To this end we use a $24 \times 16 \times 8$ $K$-grid and the PBE (GGA) functional. As a result we find distorted angles of $\alpha=$90.013$^\circ$, $\beta=$90.571$^\circ$ and $\gamma=$89.919$^\circ$, yielding a triclinic structure (changes to the lattice constants are negligible). 
This corresponds to an in-plane monoclinic distortion combined with a tilting in the direction perpendicular to the planes.
While the inter-layer geometry might suffer from neglected van-der-Waals forces, the in-plane structure is mostly governed by electron-lattice couplings which are sufficiently well captured by DFT. 
The in-plane monoclinic distortion is thus reliable and intrinsically driven already on the level of DFT.

\section{Minimal model}
\begin{figure}
\includegraphics[width=0.75\textwidth]{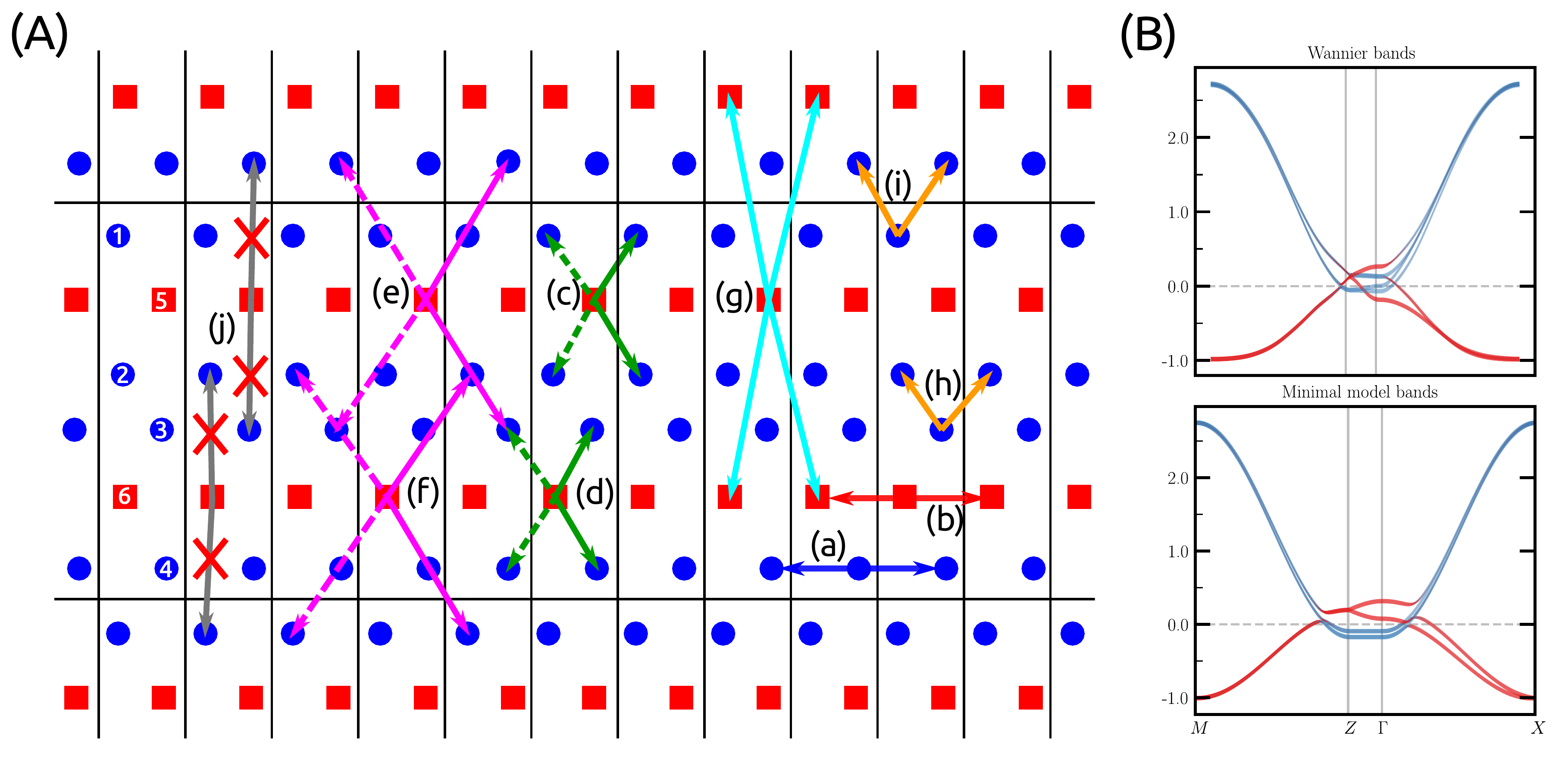}
\caption{(A) Hopping processes defining the minimal model. 
Letters are used to group different hopping processes according to the table~\ref{tab_hopping}.
(B) Comparison between the Wannier (top) and the minimal model (bottom) band structure.}
\label{fig_hop}
\end{figure}

We consider a two-dimensional minimal model with six atoms per unit cell (with one $d_{xz}$-like orbital each) reproducing the double chain structure of a Ta$_2$NiSe$_5$ layer. We take into account $i)$ single particle hoppings, $ii)$ 
intra-atomic density-density interactions, and  $iii)$ nearest-neighbor density-density interactions.
For simplicity we recall here the definition of the Hamiltonian which we already introduced in the main text:
\begin{equation}
\begin{split}
\hat{\cal H} =& \hat{H}_{hop} + \hat{H}_U + \hat{H}_V \\
=& \sum_{\vec{R} \s} \sum_{\vec{\d}} \Psi^{\dagger}_{\vec{R}+\vec{\d}} \mathbf{T}(\vec{\d}) \Psi_{\vec{R} \s} +
U \sum_{\vec{R}} \sum_{j=1,\ldots,6} 
\hat{n}_{j \up} (\vec{R}) 
\hat{n}_{j \down} (\vec{R}) 
+ \\
 &+ V \sum_{j=1,2} \sum_{\vec{R}  \s \s'} \left[ \hat{n}_{j \s}(\vec{R}) +    
\hat{n}_{j \s}(\vec{R}+\vec{\d}_x) 
\right] \hat{n}_{5 \s'}(\vec{R})  
+
V \sum_{j=3,4} \sum_{\vec{R}  \s \s'} \left[ \hat{n}_{j \s}(\vec{R}) +    
\hat{n}_{j \s}(\vec{R}-\vec{\d}_x) 
\right] \hat{n}_{6 \s'}(\vec{R}), 
\end{split}
\end{equation}
with $\Psi_{\vec{R} \s}$ a spinor defined as
$\Psi_{\vec{R} \s}^{\dagger} \equiv \left( \cc_{1\s}(\vec{R})~\cc_{2\s}(\vec{R})~\cc_{5\s}(\vec{R})~\cc_{3\s}(\vec{R})~\cc_{4\s}(\vec{R})~\cc_{6\s}(\vec{R})  \right)$
and $\hat{n}_{i \sigma}(\vec{R}) = \cc_{i \s}(\vec{R}) \ca_{i \s}(\vec{R})$.
The hopping matrix $\mathbf{T}(\vec{\d})$ contains intra-cell [$\mathbf{T}(\vec{0})$] as well
as nearest-cells terms [$\mathbf{T}(\pm a_x, \pm a_y)$] corresponding to the main contributions 
of the Wannier Hamiltonian derived above.
These matrix elements are summarized in the scheme of Fig.~\ref{fig_hop}(A), which includes
Ta-Ta (a) and Ni-Ni (b) intra-chain, Ta-Ni intra- (c)-(d) and inter-chain (e)-(f) hoppings
as well as inter-chain Ni-Ni (g) and Ta-Ta (j)-(h) hoppings. 
Dashed/full pairs of arrows indicate that in order to preserve the symmetry, these matrix elements
must be anti-symmetric under a reflection with respect to a plane perpendicular to the chains.
We have also indicated symmetry-forbidden Ta-Ni hybridization, that become non-zero upon symmetry breaking.
The matrix elements are summarized in the Table~\ref{tab_hopping}.
Fig.~\ref{fig_hop}(B) shows the comparison between the band structure of the minimal and the Wannier model.

\begin{table}[htbp]
\centering
\begin{tabular}{l|l}
\multicolumn{2}{c}{Hopping matrix elements $\mathbf{T}(\vec{\d})$}  \\ 
\hline
\multirow{2}{*}{Intra-chain Ta-Ta hopping (a)-(b)} 
& $T_{ii}(a_x,0) = T_{ii}(-a_x,0)= -0.72~\mathrm{eV} \quad i=1,\ldots,4$  \\
& $T_{ii}(0,0) = 1.35~\mathrm{eV}$ \\
\hline
\multirow{2}{*}{Intra-chain Ni-Ni hopping (a)-(b)} 
& $T_{ii}(a_x,0) = T_{ii}(-a_x,0)=  0.30~\mathrm{eV} \quad i=5,6$   \\
& $T_{ii}(0,0) = -0.36~\mathrm{eV}$ \\
\hline
\multirow{2}{*}{Intra-chain Ta-Ni hopping (c)-(d) } 
& $T_{15}(\vec{0}) = -T_{15}(a_x,0) = T_{25}(\vec{0}) = -T_{25}(a_x,0)=0.035~\mathrm{eV}$ \\ 
& $T_{36}(\vec{0}) = T_{46}(\vec{0}) = - T_{15}(\vec{0}) $ \\
\hline
\multirow{2}{*}{Inter-chain Ta-Ni hopping (e)-(f) } 
& $T_{45}(-a_x,a_y) = -T_{45}(a_x,a_y) = T_{35}(a_x,0) = -T_{35}(-a_x,0)=0.04~\mathrm{eV}$ \\ 
& $T_{26}(a_x,0) = T_{16}(a_x,-a_y) = T_{45}(a_x,a_y)  $ \\
\hline
\multirow{1}{*}{Inter-chain Ni-Ni hopping (g) } 
& $T_{65}(a_x,a_y)=T_{65}(a_x,0)=T_{65}(0,0)=T_{65}(a_x,0) = 0.030~\mathrm{eV}$   \\
\hline
\multirow{1}{*}{Inter-chain Ta-Ta hopping (h)-(i) } 
& $T_{61}(-a_x,a_y)=T_{61}(0,a_y)=T_{23}(0,0)=T_{23}(a_x,0) = 0.020~\mathrm{eV}$   \\
\hline
\end{tabular}
\caption{Elements of the hopping matrix $\mathbf{T}(\vec{\d})$. 
Matrix elements are grouped accordingly to the scheme in Fig.~\ref{fig_hop} with letters 
corresponding to the different hopping processes indicated by arrows.}
\label{tab_hopping}
\end{table}

\subsection{Hartree-Fock} 
We consider a single-particle variational wavefunction $\ket{\Psi_0}$ that allows 
for the breaking of the crystal symmetries.
The variational energy is computed by decoupling the interaction terms in the standard way:
\begin{equation}
 \braket{\Psi_0}
{\hat{n}_{j \up} (\vec{R}) 
\hat{n}_{j \down} (\vec{R}) }{\Psi_0} 
\approx  \braket{\Psi_0}
{\hat{n}_{j \up} (\vec{R}) } {\Psi_0} 
\braket{\Psi_0}{ \hat{n}_{j \down} (\vec{R}) }{\Psi_0} 
\end{equation}
and for $i \neq j$
\begin{equation}
 \braket{\Psi_0}
{\hat{n}_{j \s} (\vec{R}) 
\hat{n}_{i \s'} (\vec{R'}) }{\Psi_0} 
\approx
 \braket{\Psi_0}{\hat{n}_{j \s} (\vec{R})}{\Psi_0}
\braket{\Psi_0}{\hat{n}_{i \s'} (\vec{R'}) }{\Psi_0} 
-\d_{\s \s'} 
\braket{\Psi_0}{\cc_{j \s} (\vec{R}) \ca_{i \s'} (\vec{R'})}{\Psi_0}
\braket{\Psi_0}{\cc_{i \s'} (\vec{R'}) \ca_{j \s} (\vec{R})}{\Psi_0}.
\end{equation}
Taking the variation with respect to $\bra{\Psi_0}$ 
the HF Hamiltonian reads
\begin{equation}
\hat{H}_{HF} = 
\hat{H}_{hop} 
+
\sum_{\bk \s}
\Psi^{\dagger}_{\bk \s}  
\left(
\begin{matrix}
\hat{h}_{A} (\bk)&  0
\\ 0  & \hat{h}_{B} (\bk)
\end{matrix}
\right)
\Psi_{\bk \s}  
\end{equation}
where $\hat{h}_A(\bk)$ and $\hat{h}_B(\bk)$ are the decoupled interaction Hamiltonian for 
the A and B chain respectively. 
Specifically, accordingly to the atom labeling of Fig.~\ref{fig_hop},
\begin{equation}
\hat{h}_A(\bk) = 
\left(
\begin{matrix}
\d \e_1 & 0 & w_{15}^*(\bk)  \\
0 & \d \e_2 & w^*_{25}(\bk) \\ 
w_{15}(\bk)  & w_{25}(\bk) & \d \e_5
\end{matrix}
\right)
\qquad
\hat{h}_B(\bk) = 
\left(
\begin{matrix}
\d \e_3 & 0 & w_{36}^*(\bk)  \\
0 & \d \e_4 & w^*_{46}(\bk) \\ 
w_{46}(\bk)  & w_{46}(\bk) & \d \e_6
\end{matrix}
\right)
\end{equation}
with
\begin{equation}
\d \e_{i=1,2} = \frac{U}{2} n_{i} + 2 V n_{5} 
\quad 
\d \e_{i=3,4} = \frac{U}{2} n_{i} + 2 V n_{6} 
\quad  
\d \e_5 = \frac{U}{2}n_5 + 2V \left( n_{1} + n_2 \right)
\quad
\d \e_6 = \frac{U}{2}n_6 + 2V \left( n_{3} + n_4 \right)
\end{equation}
and
\begin{equation}
w_{i 5} = -V \Delta_{i 5}(\vec{0})(1 - e^{-i k_x a}) - V \phi_{i 5} e^{-i k_x a} \quad 
w_{i 6} = -V \Delta_{i 6}(\vec{0})(1 - e^{ i k_x a}) - V \phi_{i 6} e^{ i k_x a}.
\end{equation}
In these equations $n_{i} = \braket{\Psi_0}{\cc_{i \s}(\vec{0}) \ca_{i \s}(\vec{0})}{\Psi_0}$,
$\Delta_{ij}(\vec{0}) = \braket{\Psi_0}{\cc_{i \s} (\vec{0}) \ca_{j \s} (\vec{0})}{\Psi_0}$ and 
$\phi_{ij}$ are the order parameters defined in the main text. 
All the above parameters are self-consistently determined by diagonalizing the HF Hamiltonian 
starting from an initial guess.

\subsection{Double Counting Corrections} 

To avoid double counting of correlation effects within our Hartree-Fock calculations which are already present on the level of the DFT calculations, we make use of cRPA Coulomb matrix elements and apply a double counting correction potential to the bare band structure.  The former aims to avoid a double counting of screening processes to the Coulomb interactions resulting from the model band structure. By excluding these screening processes in cRPA calculations for the Coulomb matrix elements based on the full ab initio band structure we take screening from the "rest" of the band structure into account, but not from the bands of the minimal model. A double counting of this kind is thus avoided in the interaction terms. On the other side, the double-counting potential is introduced to not count twice the effect of local interactions already included in DFT. The commonly used ansatz for this is an orbital-independent potential which is acting only on the correlated orbitals. Since our minimal model is completely down-folded to correlated orbitals only, a potential of this form would equally shift all involved bands and would thus have no effect at all. We can thus safely neglect a double-counting correction potential of this form.

Care must, however, be taken due to the use of the modified Becke-Johnson (mbj) exchange potential (in contrast to GGA or LDA approximations). Although this exchange potential is still local, it effectively accounts for non-local Coulomb interaction terms here. The most prominent effect of the mbj exchange potential for Ta$_2$NiSe$_5$ is to decrease the overlap between the mostly Ta-like conduction bands with the mostly Ni-like valence bands, which is controlled by the mbj parameter $c_\text{mbj}$. For $c_\text{mbj} = 1.0$ the results are very similar to GGA/LDA calculations with a an overlap of about $400\,$meV at $Z$. For our self-consistently calculated $c_\text{mbj}=1.26$ the overlap is approx. $200\,$meV. In order to not double-count this decreasing trend of the overlap upon inclusion of correlation effects, the Ta and Ni onsite energies of our minimal model are adjusted to result in an overlap of about $400\,$meV (see Fig.~\ref{fig_hop} B). We also checked the influence of this procedure and find that all of our conclusions hold independently on the exact value of this change in the overlap. The phase diagram looks qualitatively the same and just slight quantitative changes are observed so that the critical values $V_l^*(U)$ and $V_u^*(U)$ shift to slightly larger values upon increasing the band overlap in the bare minimal model.

\bibliography{suppl_TNS.bib}